\documentclass[preprintnumbers,prd,showpacs,floatfix,preprintnumbers,
twocolumn,amsmath,amssymb,nofootinbib,
superscriptaddress]
{revtex4}
\usepackage{latexsym}
\usepackage{epsfig}
\usepackage{amssymb}
\usepackage{amsmath}
\usepackage{dcolumn}
\usepackage{bm}
\usepackage{graphicx}

\begin{document}

\title{
Resembling dark energy and modified gravity with Finsler-Randers cosmology}

 \author{S. Basilakos}\email{svasil@academyofathens.gr}
\affiliation{Academy of Athens, Research Center for Astronomy and
Applied Mathematics,
 Soranou Efesiou 4, 11527, Athens, Greece}

 \author{A.P.Kouretsis }
\email{akouretsis@astro.auth.gr}
\affiliation{Section of Astrophysics, Astronomy and Mechanics, Department of
Physics Aristotle University of Thessaloniki, Thessaloniki 54124, Greece}

 \author{Emmanuel N. Saridakis}
\email{Emmanuel\_Saridakis@baylor.edu}
 \affiliation{Physics Division, National Technical University of Athens,
15780 Zografou Campus,  Athens, Greece}
\affiliation{Instituto de F\'{\i}sica, Pontificia Universidad de Cat\'olica
de Valpara\'{\i}so, Casilla 4950, Valpara\'{\i}so, Chile}

\author{P.C. Stavrinos}
\email{pstavrin@math.uoa.gr}
\affiliation{Department of Mathematics, University of Athens, Athens 15784,
Greece}


\begin{abstract}
In this article we present the cosmological equivalence between the
relativistic Finsler-Randers cosmology, with dark energy and modified gravity
constructions, at the background level. Starting from a small deviation from
the quadraticity of the Riemannian geometry, through which the local
structure of General Relativity is modified and the curvature theory is
extended, we extract the modified Friedmann equation. The corresponding
extended Finsler-Randers cosmology is very interesting, and it can mimic
dark-energy and modified gravity, describing a large class of scale-factor
evolutions, from inflation to late-time acceleration, including the phantom
regime. In this respect, the non-trivial universe evolution is not attributed
to a new scalar field, or to gravitational modification, but it arises from
the  modification of the geometry itself.
\end{abstract}

\pacs{98.80.-k, 95.36.+x, 04.50.Kd}

\maketitle

\section{Introduction}

Since the discovery of the accelerated expansion of the Universe
(see \cite{Teg04}
and references therein)
a lot of effort has been made in order to understand
the physical mechanism which is responsible
for such a cosmological phenomenon. There are two basic directions
one can follow in order to obtain its explanation. The first is to introduce
the concept of dark energy (hereafter DE)  within the framework of General
Relativity (for reviews see for instance  \cite{Ame10}),
while the second is to modify the gravitational sector itself (see
\cite{Capozziello:2011et} and
references therein).

From the DE viewpoint the simplest way to fit the current cosmological data
is to include in the Friedmann equations the cosmological constant
\cite{Teg04}.
However, the disadvantage of the so-called concordance $\Lambda$-cosmology is
the fact that it suffers from the cosmological constant problem itself
\cite{Weinberg89}. This intrinsic problem appears
as a difficult issue which includes
many aspects: not only the problem of understanding the tiny current value
of the vacuum energy density ($\rho_{\Lambda}=c^{2}\,\Lambda/8\pi G\simeq
10^{-47}\,GeV^4$) \cite{Weinberg89} in the context of quantum field theory
or string theory, but also the cosmic coincidence problem, namely why
the density of matter is now so close to the vacuum density
\cite{Steinhardt97}.
Unfortunately, the alternative and more complex DE scenarios, for instance
quintessence  \cite{Caldwell98,Peebles03,Pad03}, phantom
\cite{phant}, quintom \cite{quintom} etc,
are not free from similar fine-tuning and other no less severe
problems (including the presence of extremely tiny masses and peculiar
forms of the scalar field kinetic energy).

The above problems have inspired many authors to proceed to the alternative
direction of modified gravity, such as the braneworld
Dvali, Gabadadze and Porrati \cite{Dvali2000} model,
$f(R)$ gravity \cite{Sot10}, $f(T)$ gravity \cite{Ben09,Ferraro:2006jd},
scalar-tensor theories \cite{scal}, Gauss-Bonnet gravity \cite{gauss},
Ho\v{r}ava-Lifshitz gravity \cite{hor}, nonlinear massive gravity
\cite{Hinterbichler:2011tt} etc.
The underlying idea is that the accelerated expansion, either during
inflation or at late  times, can be driven by a modification of the
Einstein-Hilbert action, while the matter content of the universe remains
the same (relativistic and cold dark matter).
However, the majority of modified gravity models are plagued with no
physical basis and/or many parameters.

On the other hand, the last decade the Finslerian relativistic extensions
 have gained a lot of attention, since
Finsler geometry naturally extends the traditional Riemannian geometry
\cite{Rad41}.  In this formulation, in general one starts with the Lorentz
symmetry breaking, which is a common feature within quantum gravity
phenomenology. Such a departure from relativistic symmetries of space-time,
leads to the possibility for the underlying physical manifold to have a
broader geometric structure than the simple pseudo-Riemann geometry. In these
lines, Finsler geometry is the simplest class of extensions, since it
generalizes Riemann geometry.
  Note that the Riemannian geometry itself is a special type of the
Finslerian one.

One of the most characteristic features of Finsler geometry is the dependence
of the metric tensor to the position coordinates of the base-manifold and to
the tangent vector of a geodesic congruence, and this velocity-dependence
reflects the Lorentz-violating character of the kinematics. Additionally,
Finsler geometry is strongly connected to the effective geometry within
anisotropic media \cite{Born} and naturally enters the analogue gravity
program \cite{Barcelo:2005fc}. These features suggest that Finsler geometry
may play an important role within quantum gravity physics.

From the cosmological viewpoint, in a series of works \cite{Stavrin,Bastav13}
it was reported that in the osculating Riemannian limit the cosmic expansion
of the flat Finsler-Randers (hereafter FR) gravity is identical to that of
flat DGP, despite the fact that the geometrical origin of the two
cosmological models is completely different. The latter means that the flat
FR model inherits all the advantages and disadvantages of the flat DGP
gravitational construction. However, the fact that DGP gravity is under
observational pressure \cite{Fair06} implies that the flat FR model faces the
same problems \cite{Bastav13}.

Therefore, in the present work we are
interested in extending the results of \cite{Stavrin} and \cite{Bastav13} in
order to derive an extended version of the FR model (hereafter EFR), free
from the observational inconsistencies. To achieve that, instead of the
osculating Riemannian limiting processes  \cite{Stavrin}, which is a
metric-based approach, we use the covariant 1+3 formalism \cite{Relcos}, that
under certain conditions can be naturally extended in the Finslerian
framework \cite{Kouretsis:2012ys}. In this less restrictive case, we can
mimic all   non-interacting  DE models and the majority of modified
gravitational constructions,
and we are able to describe a large class of cosmological evolutions.

The plan of the work is as follows. In Sec. \ref{Finslermodel} we present the
 metrical extension of Riemannian geometry, and we discuss the evolution of
the kinematical variables and the Finsler-Randers geometrical structure. In
Sec. \ref{Finslercosmology} we focus on the isotropic expansion and we
develop the cosmological model. We prove the equivalence between the EFR and
DE, as well as with some classes of modified gravity, at the expansion level,
and we discuss some particular examples. Finally, in Sec. VI we draw our
conclusions.

\section{Relativistic Finsler geometry}
\label{Finslermodel}

Recently, there is an increasing interest in Finsler geometry since
it has been
reported within different aspects of quantum gravity. The effective metric
depends either on velocity-like
variables or on the tangent vector field of the observers' cosmic lines. A
representative example of the first case is the stochastic space-time D-foam
where the effective metric depends on the velocity of D-particles that recoil
on the world-sheet \cite{Mavromatos:2010ar}. Another scenario where Finsler
geometry emerges, and the metric depends on fiber coordinates, is the
covariant
Galilean transformations in curved space-times \cite{Germani:2011bc}. On the
other hand, dependence of the metric on the particle's 4-velocity arises in
other Lorentz-violating theories,  such as the  Ho\v{r}ava-Lifshitz gravity
\cite{Romero:2009qs}. Additionally, a Finslerian line-element has been
encountered in deformations of Cohen and Glashow's very special relativity
\cite{Cohen:2006ky}, as well as in holographic fluids \cite{Leigh:2011au}.
Moreover, bi-metric constructions can be naturally incorporated in the
Finsler framework \cite{Skakala:2010hw}. Finally, we mention that Finsler
geometry can be closely related to the standard-model extension
\cite{Kostelecky:2011qz}. Before proceeding to the cosmological
application of Finsler geometry, in the following subsections we briefly
present its basic features.

\subsection{Finsler congruences}

The main object in Finsler geometry is the fundamental function $F(x,dx)$
that generalizes the Riemannian notion of distance (see for example
\cite{RundFB,Shen,Bao}). In Riemann geometry the latter is
a quadratic function with respect to the infinitesimal increments $dx^a$
between two neighboring points. Keeping all the postulates of Riemann
geometry but accepting a non-quadratic distance measure, a metric tensor
can be introduced as
\begin{equation}
g_{ab}(x,y)={1\over2}\frac{\partial^2F^2}{\partial y^a\partial
y^b},\;\;\;\;y^a\neq0,
\label{mete}
\end{equation}
for a given connecting curve with tangent $y^a={dx^a\over d\tau}$. Note that
when the generating function $F(x,y)$ is quadratic, the above
definition is still valid and leads to the metric tensor of Riemann geometry.
The
dependence of the
metric tensor to the position coordinates $x^a$ and to the fiber coordinates
$y^a$ suggests that the geometry of Finsler spaces is a geometry on the
tangent bundle $TM$. In other words, the Finsler manifold is a fiber space
where tensor fields depend on the position and on the infinitesimal
coordinate increments $y^a$. Therefore, the position dependence of  Riemann
geometry is replaced by the so called {\it element of support}, which is the
pair $(x^a,y^a)$.

In relativistic applications of Finsler geometry the role of the supporting
direction $y^a$ must be explicitly given. For example, it may stand
as an internal variable, as an explicit or implicit violation of Lorentz
symmetry, as an aether-like direction or simply as the velocity of the
fundamental observer. In this article we restrain our analysis to the latter
case, where the supporting direction $y^a$ is the tangent to the cosmic flow
lines. Using only variational arguments we can arrive to the deviation
equation for the supporting congruence $y^a$. The deviation equation directly
provides all the information  for the internal deformation of the time-like
geodesic flow $y^a$. Following the same procedure with GR, we can extract the
propagation formulas for the expansion, shear and vorticity of an
infinitesimal cross-section of the cosmological flow.

The infinitesimal distance between two neighboring points on the base
manifold (position space) is given by the small displacement along the
connecting curve $\gamma(\tau)$, that depends on the position $x^a$ and on
the
coordinate increments $dx^a$:
\begin{equation}
d\tau=F(x,dx),
\end{equation}
where in Finsler geometry $F(x,dx)$ does not necessarily
depend quadratically on the $dx^a$ increments. The actual distance traveled
on the base
manifold along a given direction is
\begin{equation}
I=\int F(x,dx/d\tau)d\tau,
\end{equation}
where the metric function $F(x,y)$ is homogeneous of first order with respect
to the displacement arguments $y^a=dx^a/d\tau$. Applying the least action
principle on the previous integral we arrive to the geodesic equation for the
supporting direction
\begin{equation}
\dot{y}^a\equiv y^a\nabla_a y^b= {dy^a\over d\tau}+2G^a(x,y)=0,\label{gede}
\end{equation}
where $G^a$ are the spray coefficients with respect to the $F(x,y)$
fundamental function and they are given by
\begin{equation}
G^a={1\over4}g^{ab}\left({\partial^2F^2\over \partial x^c \partial
y^b}y^c-{\partial F^2\over \partial x^b}\right)\label{spr}.
\end{equation}
When $y^a$ stands for the velocity of the fundamental observer then the dot
operator in relation (\ref{gede}) is a direct generalization of the
time-propagation of GR relativistic kinematics. In other words, if the
supporting direction $y^a={dx^a\over d\tau}$  is the observers' 4-velocity
then the affine parameter $\tau$ is the proper time. Note, that we can recast
relation (\ref{gede}) in the familiar form with respect to the Christofell
symbols, with the only difference that the metric will depend on the
supporting element.

The relative acceleration between two neighboring observers is given by the
second variation of the distance module. Focusing the analysis along the
$y^a$ direction, the second variation leads to the Jacobi equation. The
relative acceleration between nearby geodesics is monitored by an
infinitesimal connecting vector (the deviation vector) defined as
\begin{equation}
\tilde{x}^a=x^a+\xi^a,
\end{equation}
where the tilde stands for the neighboring reference frame. Then,
substituting  the previous expression to the Euler-Lagrange equations
(\ref{gede}), and keeping up to first-order terms with respect to the
deviation vector $\xi^a$, leads to the following formula
\begin{equation}
\ddot{\xi}^a+\mathcal{H}^a_{\;\;b}(x,y)\xi^b=0,\label{deviat}
\end{equation}
where $\mathcal{H}^a_{\;\;b}$ is a tensor field that incorporates the
relative
displacement of nearby geodesics in a Finslerian framework, given by
\begin{equation}
\mathcal{H}^a_{\;\;b}=2\frac{\partial G^a}{\partial
x^b}-y^c\frac{\partial^2G^a}{\partial y^b \partial
x^c}+2G^c\frac{\partial^2G^a}{\partial y^b\partial y^c}-\frac{\partial
G^a}{\partial y^c}\frac{\partial G^c}{\partial y^b}.\label{devte}
\end{equation}
The first order homogeneity of the metric function
leads to the constraint ${\partial g_{ab}\over\partial y^c}y^c=0$. The latter guarantees that for most connection  structures (for example Chern, Cartan or
Berwald) the Jacobi field (\ref{deviat}) remains the same (see for example \cite{RundFB,Bao,Shen,Miron}).

The coefficients of the tensor field $\mathcal{H}^a_{\;\;b}$ are directly
determined by the metric function $F(x,y)$ through the least action principle
that gives back the spray coefficients (\ref{spr}). As in Riemann geometry,
expression (\ref{devte}) is second order homogeneous with respect to $y^a$,
but the dependence is non-quadratic. It's eigenvalues correspond to the
sectional curvatures in the principal directions and designate the relative
motion between neighboring integral curves.  Relation (\ref{devte})  encloses
all the relevant information for Finslerian tidal effects on the $y^a$
congruence. The tensor field $\mathcal{H}^a_{\;\;b}$ is responsible for the
relative acceleration between nearby observers and will generate expansion
and shear on the time-like $y^a$-congruence. Apparently, the $y^a$-deformable
kinematics will be modified due to the non-quadratic dependence of
$\mathcal{H}^a_{\;\;b}$ on the velocity of the fundamental observer.

\subsection{Deformable kinematics}
\label{kinematics}

The observers' time-like congruence introduces a uni-direction in the
physical manifold. This asymmetry is encoded in the metric function $F(x,y)$
and induces the $1+3$ ``threading'' of space-time \cite{Relcos}. In the
covariant 1+3 formalism the metric is not the central object, since we do not
use a particular coordinate system. Instead, we use the kinematic quantities,
the irreducible components of curvature and conservation arguments, while
Einstein's field equations enter as simple algebraic relations between
curvature and matter \cite{Relcos}. The deviation of geodesics is of central
importance since it monitors the internal deformation of the cosmic medium in
a covariant way.

 From the Finslerian perspective
the space and time decomposition  is directly related to the first-order
homogeneity of $F(x,y)$. In particular, the fundamental observer's velocity
$y^a$ defines a family of integral curves on the space-time manifold.  With
respect to this 4-velocity we can decompose tensor fields along $y^a$ and on
the perpendicular  spatial hyper-surface. In fact, we can recast the metric
tensor (\ref{mete}) in the following form
\begin{equation}
g_{ab}=F\frac{\partial^2F}{\partial y^a \partial y^b}+\frac{\partial
F}{\partial y^a}\frac{\partial F}{\partial y^b},\label{spl}
\end{equation}
where we have split the space-time metric in two parts by using the
quantities
\begin{equation}
l_a=\frac{\partial F}{\partial
y^a}\;,\;\;\;\;h_{ab}=F\frac{\partial^2F}{\partial y^a\partial
y^b}.\label{dec}
\end{equation}
Using the first order homogeneity of the metric function $F(x,y)$ we can
prove that $l_a$ is the normalized velocity of the observers' flow-lines,
$l_a=y_a/F$. In addition, the first order homogeneity of the fundamental
function implies that $h_{ab}l^b=0$ and also that the rank is $(h_{ab})=3$.
Therefore, the tensor $h_{ab}$ stands for the projection tensor of
relativistic kinematics.

With the space-time split (\ref{spl}) in hand we can decompose tensor fields
to their irreducible parts, in direct analogy to the standard gravitational
physics, for example
\begin{equation}
X_a=g_a^{\;\;b}X_b=(h_a^{\;\;b}+l^bl_a)X_b=Xl_a+\mathcal{X}_a,
\end{equation}
where $X=l_aX^a$ is the time-like part and $\mathcal{X}_a=h_a^{\;\;b}X_b$ is
the space-like part. Using the $1+3$ covariant formalism we will track the
internal motion of the normalized supporting direction $l^a$ (the congruence
$l^a$ is considered to be time-like, $l_al^a=1$ \cite{Stavrinos2005}).
Restraining the analysis
along the $l$-time-like flow, the propagation equation  of the deviation
vector at first order is given by
\begin{equation}
\dot{\xi}^a=B^a_{\;\;b}\xi^b,\label{fde}
\end{equation}
where it is straightforward to prove that $B^a_{\;\;b}=\nabla_bl^a$, that
is the
tensor field
$B^a_{\;\;b}$ is the distortion tensor of the time-like congruence.
Following relation (\ref{dec}), we can decompose the distortion tensor to
it's irreducible parts
\begin{equation}
\nabla_bl_a={1\over3}\Theta h_{ab}+\sigma_{ab}+\epsilon_{abc}\omega^c,
\end{equation}
where for the 3D spatial derivative ${\rm D}_a=h_a^{\;\;b}\nabla_b$ the
irreducible components  are: the expansion $\Theta={\rm D}^al_a$ that tracks
volume changes, the shear $\sigma_{ab}={\rm D}_{\langle
b}l_{a\rangle}$\footnote{Angle brackets stand for the projective, symmetric
and   trace-free part  of a second rank tensor  $X_{\langle
ab\rangle}=h^c_{\;\;(a}h^d_{\;\;b)}X_{cd}-{1\over3}X_{cd}h^{cd}h_{ab}$. } that
incorporates shape distortions, and the vorticity
$\omega_a=\epsilon_{abc}{\rm
D}^bl^c/2$ that accounts for changes of the orientation of the infinitesimal
spatial cross-section, parallel transported along $l^a$.

Taking the time derivative of relation (\ref{fde}) and substituting in
relation (\ref{deviat}) we arrive to the evolution equation for the internal
deformations of the time-like flow:
\begin{equation}
\dot{B}_{ab}+B_{ac}B^{c}_{\;\;b}=-\mathcal{H}_{ab}.\label{defev}
\end{equation}
This propagation law reflects the effect of the Finslerian curvature
tensor $\mathcal{H}_{ab}$ on the deformable kinematics of a time-like flow.
The irreducible parts of relation (\ref{defev}) provide the evolution
equation for the expansion (Raychaudhuri's equation)
\begin{equation}
\dot\Theta+{1\over3}\Theta^2=-\mathcal{K}-2(\sigma^2-\omega^2),\label{exp}
\end{equation}
the propagation of shear (which describes kinematic anisotropies)
\begin{equation}
\dot{\sigma}_{\langle
ab\rangle}=-\frac{2}{3}\Theta\sigma_{ab}-\sigma_{c\langle
a}\sigma^c_{\;\;b\rangle}-\omega_{\langle
a}\omega_{b\rangle}-\mathcal{H}_{\langle ab\rangle},\label{she}
\end{equation}
 and finally the propagation of
vorticity
\begin{equation}
\dot{\omega}_a=-{2\over3}\Theta\omega_a+\sigma_{ab}\omega^b,\label{vor}
\end{equation}
where $\mathcal{K}=\mathcal{H}_{ab}h^{ab}$ is the scalar flag curvature of
the Finslerian manifold when $\mathcal{H}_{\langle ab\rangle}=0$
 \cite{Shen,Kouretsis:2012ys}.

The above system of propagation equations is very
similar to the analogous expressions in General Relativity. In particular,
they
are the same except that in the Riemannian limit the tensor (\ref{devte})
depends quadratically on the observers' 4-velocity. Thus, the key
difference is
the non-trivial dependence
of the curvature tensor $\mathcal{H}_{ab}$ to the velocity of the fundamental
observer, which modifies the way that curvature generates deformations on a
time-like medium.

\subsection{Finsler-Randers metric function}

In Finsler geometry the form of the metric function $F(x,y)$ is of central
importance since it generates all the other geometric quantities. One of the
most simple cases after the Riemann limit is the Randers norm \cite{Ran},
which is given by
\begin{equation}
F=\alpha+\beta,\label{ran}
\end{equation}
where $\alpha=\sqrt{\alpha_{ab}y^ay^b}$ is a Riemann metric function and
$\beta=b_ay^a$ stands for an arbitrary 1-form. The fundamental function
(\ref{ran}) interfaces a Riemann space-time with a Finslerian one in a simple
way, since the Randers metric is the limiting case of a large number of
Finsler space-times, when we consider small departures from GR. For example,
in a
large class of $(\alpha,\beta)$-metrics where $F=\alpha\phi(\beta/\alpha)$,
the almost Riemannian limit $\phi\sim1$ provides a Randers type geometry at
 first-order   for $\beta/\alpha$, when $\phi\sim 1+\beta/\alpha$.

Relation (\ref{ran}) has the important consequence that one can separate
geometric quantities to the purely Riemann part
with respect to the $\alpha$ metric function, and to the
Finsler contribution. In this
case, for specific examples, we can directly inspect the effect of
non-quadraticity on the space-time medium.

In particular, the geometric entity that accurately incorporates the
non-quadraticity of the metric function is the indicatrix $F(x,y)=1$, which
represents an
arbitrary locus on the tangent bundle \cite{RundFB}. This locus in the
Riemann case defines a quadratic hyper-surface. In case of a Randers type
geometry (\ref{ran}) the hyper-surface is still quadratic but becomes
eccentric \cite{Ran}. In other words,
the Randers metric function (\ref{ran}) assigns at each space-time point a
vector $b_a$ that describes the displacement of the center  of the
indicatrix.
This property translates to a disformal correlation between the Finslerian
metric tensor (\ref{mete}) and the Riemannian $a_{ab}$ given in relation
(\ref{ran}). In fact, substitution of relation (\ref{ran}) into (\ref{mete})
yields
\begin{equation}
 g_{ab}={F\over\alpha}(\alpha_{ab}-\bar l_a\bar l_b)+l_a l_b,
 \label{Rmet}
\end{equation}
where $\bar l_a={\partial \alpha\over \partial y^a}$ is the normalized
velocity on the Riemann sector. The disformal relation (\ref{Rmet})
introduces an explicit dependence of the space-time metric on the velocity of
the cosmic flow lines. Note, that similar behavior of the effective geometry
is commonly reported in investigations of anisotropic media \cite{Born}.

  Concerning the signature of the Finsler-Randers space-time it is useful to
introduce
the non-holonomic frame on $TM$, namely
\begin{equation}
Y_a^{\;\;b}=\sqrt{{\alpha\over F}}\left(\delta_a^{\;\;b}+\sqrt{\alpha \over F}l_al^b-\bar l^a\bar l_b\right) \label{nonfr}
\end{equation}
and it is straightforward to prove the identity $Y_a^{\;\;c}Y_c^{\;\;b}=\delta_a^{\;\;b}$. Then, using relation
(\ref{nonfr}) we can recast the Finslerian metric tensor to the following form \cite{Holland:1982bf}
\begin{equation}
g_{ab}=Y_a^{\;\;c}Y_b^{\;\;d}\alpha_{cd}.
\end{equation}
Thus, from the definition (\ref{nonfr}) and the above relation we conclude
that the Finsler-Randers metric tensor and the Riemann metric $\alpha_{ab}$
have the same signature.  Taking into account that the metric tensor
(\ref{Rmet}) must be real, we deduce that the time-like $y^a$-congruence is
positive definite or equivalently
the signature is $(+,-,-,-)$. This restrains the $y^a$-bundle of geodesics to
be time-like \cite{Skakala:2010hw}, however we can define first-order
Finslerian tensor fields as space-like $U(x,y)^aU(x,y)_a<0$,
null $U(x,y)^aU(x,y)_a=0$, and time-like $U(x,y)^aU(x,y)_a>0$ \cite{Beam}.

As a first attempt to examine the physical impact of non-quadraticity, we
assume that the Riemann sector of relation (\ref{Rmet}) represents the
``gravitational'' geometry, while the Finsler manifold describes the
``physical'' geometry \cite{Bekenstein}. This provides an effective
geometric setup to model possible implications of non-quadraticity on the
expansion dynamics.

We consider a Randers type $(\alpha,\beta)$-metric and we assume that the
velocity
of the fundamental observer is given by the normalized vector $l^a=y^a/F$.
Then, if $b_a$ is a closed form with respect to the Riemann covariant
derivative of the $\alpha$-metric, $b_{[a;b]}=0$, the spray coefficients
(\ref{spr}) for the normalized velocity $l^a$  take the simplified form
\cite{Shen}
\begin{equation}
G^a=\bar G^{\;a}+{1\over2}\Phi l^a,\label{ransp}
\end{equation}
where   bars denote  the Riemann parts with respect to the $\alpha$-metric,
 and we define $\Phi=b_{a;b}l^al^b$. Substituting relation
(\ref{ransp}) into the covariant expression (\ref{devte}), the curvature
tensor
takes the simplified form
\begin{equation}
\mathcal{H}^a_{\;\;b}=\bar {\mathcal
H}^a_{\;\;b}+{1\over4}\left(3\Phi^2-2\Psi\right)h^a_{\;\;b},\label{Hran}
\end{equation}
where  the last two scalars are given
with respect to the Riemann covariant derivative of the $\alpha$-metric:
\begin{equation}
\label{PhiPsi}
\Psi=b_{a;b;c}l^al^bl^c,
\end{equation}
and we have defined $\bar {\mathcal H}_{ab}=F^{-2}\bar R_{acbd}y^cy^d$ for
the
part of the
curvature coming from the Riemann metric function $\alpha$
of relation (\ref{ran}).
The second rank tensor (\ref{Hran}) clearly incorporates the relation
between a part of the
Riemann curvature and a part the actual curvature of the foreground manifold.
As we have already mentioned, the latter curvature generates deformations in
the assumed ``physical'' space-time
which is of Finsler type, while the Riemann curvature represents the
gravitational sector. Note that $\Psi$, that is
the nature of the Finsler-Randers contribution to the curvature, is
defined on a geometrical basis, since it directly
originates from the curvature of the Finsler-Randers geometry (\ref{Hran}).
Therefore, from relation (\ref{Hran}) and the
deformable kinematics given in relations (\ref{exp})-(\ref{vor}), we conclude
that the Riemann curvature of the gravitational sector generates deformations
in the foreground space-time in a modified way.

\section{ Finsler-Randers cosmology}
\label{Finslercosmology}

In this section  we investigate the conditions under which the
Finsler-Randers
cosmology can provide a cosmic acceleration equivalent to the traditional
scalar field DE or classes of modified gravity. We assume that the
``physical'' geometry is represented by the non-quadratic
metric function (\ref{ran}), while the gravitational geometry is given by
its' Riemannian part.  Thus, in a FRW-like scenario the Riemann curvature
$\bar R_{abcd}$ is related to the energy-momentum tensor of a perfect fluid through
the Einstein's field equations
\begin{eqnarray}\nonumber
\bar R_{ab}-{1\over2}\bar R\alpha_{ab}&=&T_{ab}\\
&=&\rho\bar l_a\bar l_b+ p\bar h_{ab}, \label{feq}
\end{eqnarray}
where we define the projection tensor of the Riemannian sector as
$\bar h_{ab}\equiv\alpha_{ab}-\bar l_a \bar l_b$,
the overall energy density as measured in the $\bar l^a$
frame as $\rho=T_{ab}\bar l^a\bar l^b$, the total isotropic
pressure as $ p=T_{ab}\bar h^{ab}/3$ while we
impose the usual convenient units setting $8 \pi G\equiv 1$.
 Furthermore, for our setup it is natural to assume a homogeneous  and isotropic Riemannian sector
for the gravitational geometry. Hence, we can neglect the non-local gravitational
degrees of freedom and the Weyl curvature becomes negligible. In this case
the Riemann curvature depends only on its' local parts
 \begin{eqnarray}
  \bar R_{abcd}={1\over2}(\alpha_{ac}\bar R_{bd}+\alpha_{bd}\bar R_{ac}-\alpha_{bc}\bar
R_{ad}-\alpha_{ad}\bar R_{bc})\nonumber\\
-{1\over6}\bar
R(\alpha_{ac}\alpha_{bd}-\alpha_{ad}\alpha_{bc}).
\label{lri}
 \end{eqnarray}

We consider shear and vorticity free evolution for the cosmic fluid, which
is in
agreement with the tight constrains of the cosmic microwave
background (CMB)  anisotropies (see for instance \cite{ellispr}). In fact,
using relations (\ref{Hran}),(\ref{feq}) and (\ref{lri})
we obtain that $\mathcal{H}_{\langle
ab\rangle}\sim b_{\langle a}b_{b\rangle}$, and the source term in the
propagation of shear (\ref{she}) is negligible if $b_a$ tends to be purely
time-like. Thus, our kinematical setup is consistent with a shear and vorticity
free bulk flow, since there are no source terms in relations (\ref{she}) and
(\ref{vor}).    Then,
keeping up to first-order terms with respect to $b_a$ in (\ref{Hran}), and
using the field equations (\ref{feq}) and the decomposition (\ref{lri}),
Raychaudhuri's formula (\ref{exp}) acquires the simplified form

 \begin{equation}
 \dot{\Theta}+{1\over3}\Theta^2=-{1\over2}(1-\beta)(\rho+3p)-{3\over2}\Psi,
 \label{Rayfin}
 \end{equation}
where we have used the auxiliary relation $\bar
{\mathcal{H}}_{ab}h^{ab}={1\over2}(1-\beta)(\rho+3p)$.
The Raychaudhuri's equation (\ref{Rayfin}) is the fundamental equation that
describes the cosmological evolution. The crucial point is the sign of its
right hand side. In particular, negative terms align with the gravitational
pull, while positive terms accelerate the expansion.
The Finsler contribution
in the first term of the rhs of (\ref{Rayfin}) acts as an effective coupling
constant. As an example, if we neglect this term
then the significant term
that incorporates the effects of non-quadraticity is the last one, and when
$\Psi<0$ it can drive an
accelerating phase, while for $\Psi>0$ it increases the gravitational
attraction while for $\Psi=0$ the current scenario reduces to
the Einstein de-Sitter model in the matter era.
In other words, the adoption of a non-quadratic
measure affects the local structure of space-time, since the $SO(4)$ symmetry
is broken, and hence it implies new kinematic effects for the bulk flow of
matter, by modifying the curvature theory. Apparently, by discarding the
local flatness of General Relativity we acquire long-range modifications in
the Finslerian geometrodynamics.

 Let us discuss here the energy conservation in the scenario at hand.
The energy density and the isotropic pressure as measured in the Riemannian
frame $\bar l^a$ are related to the
``physical'' frame $l^a$ by the following relations
\begin{equation}
\rho={F^2\over\alpha^2}\rho^{(f)}\;\;,\;\;\;\;p={F\over\alpha}p^{(f)}
\label{frame},
\end{equation}
where we have defined $\rho^{(f)}=T_{ab} l^a l^b$ and
$p^{(f)}=T_{ab}h^{ab}/3$ for the total energy density and pressure
respectively in the Finslerian frame.
 Hence, taking into account that at late times $F/\alpha=1+\beta/\alpha\sim1$, we
obtain that at first order in the two frames the energy density and
pressure are the same, namely
 $\rho\sim\rho^{(f)}$ and $p\sim p^{(f)}$ \footnote{In relativistic cosmology
a  similar limiting process between relative frames
  is  used to study ``peculiar'' frames and the Zeldovich approximation (see
for example \cite{Zeldovich}).}.
At early times our first-order approximation is no-longer valid since the two
frames will start to diverge, having a direct impact
on the effective equation of state. Additionally, in the presence of
pressure the spatial part of the energy-momentum conservation
$h_a^{\;\;c}\nabla^bT_{cb}=0$  yields
\begin{equation}
(\rho+p)l^b\nabla_bl_a=-{\rm D}_ap.
\end{equation}
However, by construction the $l^a$-congruence is geodesic (\ref{gede}) and
the previous relation implies that ${\rm D}_ap=0$.
The latter condition is valid in an isotropic and homogeneous background, but
considering cosmological perturbations
non-geodesic congruences will be involved in the calculations (for an 1+3
treatment of non-geodesic flows see \cite{Kouretsis:2012ys}).
 Hence, our model is consistent at late times of the cosmological history
(for example at dust and radiation dominated eras) and for vanishing
gradients of pressure.

On the other hand, taking the time-like part of the energy momentum
conservation,  $l^a\nabla^bT_{ab}=0$, and decomposing it to the irreducible
parts with respect to the $l^a$-congruence, we obtain
\begin{equation}
\dot\rho=-\Theta(\rho+p)\label{cont}
\end{equation}
for the total energy density. Here we mention that the
above relation is valid for the first-order approximation, where the energy
density
is almost the same in the Remannian frame $\bar l^a$ and in the Finslerian
one $l^a$ that represents the bulk flow of matter. Introducing the
characteristic
length scale $a$ ({\it scale factor}) of the spatial volume by $dV\propto
a^3$,  we extract that for the expansion we have $\Theta=(dV)\dot{}/dV=3\dot
a/a$.
Using this expression we can recast Raychaudhuri's formula (\ref{Rayfin}) in
terms of the scale factor for late times of the
cosmological evolution as
\begin{equation}
3{\ddot a\over a}=-{1\over2}(\rho+3p)-{3\over2}\Psi\label{Raypr},
\end{equation}
where the total matter fluid itself is in general a
mixture of relativistic matter
(i.e. radiation, $\rho_{r}$ with $p_{r}=\rho_{r}/3$ )
and nonrelativistic matter (i.e. cold matter, $\rho_{m}$ with $p_{m}=0$)
components, implying $\rho=\rho_{m}+\rho_{r}$ and $p=p_{m}+p_{r}=\rho_{r}/3$.
Now using the continuity equation (\ref{cont}) together with
the Raychaudhuri's formula (\ref{Raypr}), we retrieve the modified Friedmann
equation:
\begin{equation}
\label{H2DE}
H^2(a)={1\over3}\rho-a^{-2}\int a\Psi(a) da-{C_1\over a^2}.
\end{equation}
In the above expression  $C_1$ is an integration constant, which in the
FRW limit coincides with the spatial curvature, and thus without loss of
generality in the following we set it to zero.
For the rest of our analysis we focus on the matter dominated era
(well after radiation-matter equality) in which the radiation
component is considered negligible and thus we use $\rho\equiv\rho_m$.

Equation (\ref{H2DE}) incorporates the effects of  Finsler-Randers geometry
in the expansion of the universe. We remind that $\Psi(a)$, which is
the nature of the Finsler-Randers contribution to the curvature, is
defined on a geometrical basis, since it directly
originates from the curvature of the Finsler-Randers geometry (\ref{Hran}).
 Since from first principles the evolution of $\Psi$ remains
unconstrained (this could be achieved by relating the Finsler structure to a
particular Quantum Gravity scenario, which lies beyond the scope of the
present work) any $\Psi(a)$ profile is possible. Thus, from  (\ref{H2DE})
on can deduce that a large class of scale-factor evolution can be realized
within the context of  Finsler-Randers geometry.

Let us examine the condition of a local small
departure from quadraticity, in relation  to the accelerated cosmological
expansion. In a first approach we may write
\begin{equation}
\Psi=b_{a;b;c}l^al^bl^c\sim{\beta\over\lambda_{\mathcal
F}^2},\label{qpsi}
\end{equation}
where $\lambda_{\mathcal F}$ is a characteristic length scale related to the
variation of $\beta$.
A small value of $\beta$ corresponds to a sort length
scale of the modification.
Using the approximation  $\int_{a} u\Psi(u) du\sim
\Psi a^2$, together with (\ref{qpsi}), the Friedmann equation
(\ref{H2DE}) for an accelerated phase leads to the approximate
relation
\begin{equation}
|\beta|\sim\left({\lambda_{\mathcal F}\over\lambda_H}\right)^2,\label{bes}
\end{equation}
where $\lambda_H=H^{-1}$ is the Hubble horizon. The above relation is a
rough estimation of the $\beta$-parameter, with respect to the Hubble
horizon, in order to obtain an accelerated expansion. Thus, the length
$\lambda_{\mathcal F}$ represents a characteristic scale above which the
gravitational physics is affected.
The condition $|\beta|\ll1$ can be easily
fulfilled
if $\lambda_{\mathcal F}$  is  some orders of magnitude bellow the Hubble
horizon. For example, if we assume that the gravitational sector is modified
above galactic scales (kpc) and taking into account that
$\lambda_H\sim10^{10}$pc, relation (\ref{bes}) leads roughly to
$|\beta|\sim10^{-14}$. Hence, interestingly enough, even small departures from
Lorentz invariance can lead the cosmic flow to the accelerated phase.

 Let us make a comment here on the Lorentz invariance violation. The
effective geometric formulation of the present work stands for the
geometry of space-time as measured by the comoving observers of the self
gravitating cosmic medium. Thus, the $\beta$ variable parameterizes possible
departures from Lorentz invariance in the gravitational sector. The most
stringent constraints of Lorentz violation in the gravitational sector arise
from parametrized post-Newtonian (PPN) analysis using
solar system data \cite{Will:2005va}, and the most recent results from
Gravity Probe B put an upper bound at $10^{-7}$ \cite{Bailey:2013oda}.
Therefore, the above representative example lies far inside this window.
Note that during the last years, the PPN analysis in Finsler geometry has been
developed in Ref.\cite{Lammerzahl:2012kw}, however to the best of our knowledge
the metric functions that have been used are not of Randers type.
Furthermore, the study of Lorentz violation in the gravity sector involves
possible future detection of gravitational waves
and possible Lorentz
violation corrections (see for example \cite{StavGW}), constraints on the inverse square law and
gravitomagnetic effects,
CMB anisotropies and black hole physics \cite{Kostelecky:2003fs}.
This detailed analysis eventually  will also constrain the 'Finslerity'
of the gravitational sector but lies beyond the scope of this work.

\subsection{Analogue to dark energy and modified gravity}
In this subsection we show that the above Finsler-Randers-modified Friedmann
equation (\ref{H2DE}), can mimic any dark energy scenario, through a
specific reconstruction of the $\Psi(a)$, that is of the Finsler-Randers
contribution to the curvature. For this shake we write
\begin{eqnarray}
\label{Psia}
&&\Psi(a)=\Psi_{0} X(a)/3,\;\;\;\Psi_{0}<0\\
&&H(a)=H_{0}E_{EFR}(a),
\end{eqnarray}
and
using also
  that
$\rho_{m}=\rho_{m0} a^{-3}$ the Friedmann equation  (\ref{H2DE}) writes as
\begin{equation}
\label{H2DE2}
E^2_{EFR}(a)=\Omega_{m0}a^{-3}+\Omega_{\Psi_{0}}a^{-2} Q(a).
\end{equation}
In this expression we have defined
\begin{equation}
\label{H22}
Q(a)=\int_{0}^{a} uX(u) du,
\end{equation}
while the density parameters read as $\Omega_{m0}= \rho_{m0}/3H_{0}^{2}$
and $\Omega_{\Psi_{0}}= -\Psi_{0}/3H_{0}^{2}$, with
$\Omega_{m0}+\Omega_{\Psi_{0}}=1$.\footnote{Practically, defining
$\Omega_{m0}$ 
and $\Omega_{r0}$  
as the standard nonrelativistic and radiation density parameters at the
present time, we can have
that the complete Hubble function reads as
$E^2_{EFR}(a)=\Omega_{m0}a^{-3}+\Omega_{r0}a^{-4}+\Omega_{\Psi_{0}}a^{-2} Q(a)$
in the limit of $F/\alpha=1+\beta/\alpha\sim1$ (see section III).
Note, that at the last scattering surface ($z_{CMB}$) the Hubble horizon
is $\lambda_{H}\sim 2.5\times 10^{5}pc$ which implies
that $|\beta|\ll 1$ [see Eq.(\ref{bes})]. Finally, as usual,
the above density parameters satisfy the extended sum
rule $\Omega_{m0}+\Omega_{r0}+\Omega_{\Psi_{0}}=1$.} We mention that
for mathematical convenience
$Q(a)$ is normalized to unity
at the present time.

Now we can return to the aforementioned basic question: {\it Under
  which circumstances equation (\ref{H2DE2}) can resemble that
of   dark energy}? In order to address this crucial
question we need to calculate the effective equation-of-state parameter
(hereafter EoS)
$w(a)$ for the EFR cosmology
introduced above.
 We proceed as though we would not know that the original
Hubble function is the one given by equation (\ref{H2DE2}) and we assume
that it behaves according to the typical expansion rate of the universe
where the DE is caused by a scalar field with negative pressure, namely
$P_{D}=w(a)\rho_{D}(a)$.
Therefore, for homogeneous and
isotropic cosmologies, driven by
non relativistic matter and a scalar field DE,
the first Friedmann equation is given by
\begin{equation}
\label{eq:Hgeneral}
E^{2}_{DE}(a)=\left[ \Omega_{m0}a^{-3}+\Omega_{DE0}f(a)\right]
\end{equation}
with
\begin{equation}
\label{FFA}
f(a)={\rm exp}\left\{ -3\int_{1}^{a} \left[
  \frac{1+w(u)}{u}\right] {\rm d}u \right\} \;\;,
\end{equation}
where $\Omega_{DE0}= \rho_{DE0}/3H_{0}^{2}$
is the DE density parameter at   present time, which obeys
$\Omega_{m0}+\Omega_{DE0}=1$.

The next step is to require the equality of the expansion
rates of the original EFR
picture (\ref{H2DE2})  and that of
the DE picture (\ref{eq:Hgeneral}), namely $E_{RF}(a)=E_{DE}(a)$ for every
scale factor, and doing so we extract the integral equation
\begin{equation}
\label{FFA1}
Q(a)=a^{2}f(a)\,.
\end{equation}
Differentiating the above equation, and using  (\ref{H22}) and (\ref{FFA}),
we obtain the function $X(a)$ [and thus $\Psi(a)$] in terms of the
EoS parameter $w(a)$, as
\begin{equation}
\label{FFA2}
X(a)=-\left[1+3w(a)\right]\;f(a) \,.
\end{equation}
 In this viewpoint,
if we know apriori the effective EoS parameter then we can obtain
via Eq.(\ref{FFA2}) the Finser-Randers function $X(a)$ and vice-versa.
Finally, inverting (\ref{FFA2}) and utilizing again  (\ref{H22}),(\ref{FFA}),
we find after some simple algebra that
\begin{equation}
\label{FFA3}
w(a)=-1-\frac{a}{3}\left[ -\frac{2}{a}+\frac{d\ln Q}{da}\right] \;.
\end{equation}

Relation (\ref{FFA3}) is one of the basic results of our work. It provides
the relation of any EoS evolution with the necessary form of the
Finsler-Randers
geometry. In particular, for a given desired form of $w(a)$ we use
(\ref{FFA3}) in order to find the corresponding $Q(a)$.
Then through
(\ref{FFA1}),(\ref{FFA2})  we calculate $X(a)$, and using (\ref{Psia}) we
obtain $\Psi(a)$. Finally, with the profile of $\Psi(a)$ in hand we determine
the Friedmann equation of motion (\ref{H2DE})  that together with the
continuity equation (\ref{cont}) fully determines the cosmological evolution.

We stress here that there is not any restriction at all, namely the above
procedure can be applied for any $w(a)$, as long as the corresponding Hubble
function is given by  (\ref{eq:Hgeneral}), for instance including the
quintessence and phantom regimes, the phantom-divide crossing from both
sides, etc. In the following subsection, without loss of generality, we
reconstruct $X(a)$ of the Finsler-Randers metric function, for the most
  familiar  cosmological scenarios.

\subsection{Specific examples}

In order to proceed to specific examples, the precise functional form
of $X(a)$ has to be determined. However, note that this is also the case for
any dark-energy model, as
far as the equation of state (EoS) parameter in concerned.
Potentially, in the current work we could
phenomenologically treated $X(a)$ [and thus $Q(a)$ and $\Psi(a)$]
either as a Taylor expansion
around $a=1$ [$X(a)=X_{0}+X_{1}(1-a)$] or as a power law
$X(a)\propto a^{\nu}$. Instead of doing that we
have decided to mathematically investigate the conditions under which
the Finsler-Randers cosmological
model can produce some of the well known DE models.
Bellow we provide some specific examples along the above lines.
In particular, we first consider some literature 
scalar-field DE models that emerge from
FRW cosmology with General Relativity, and for these models we reconstruct
the
functional forms of $\Psi(a)=\Psi_{0}X(a)/3$ of the equivalent
Finsler-Randers cosmology.

\begin{itemize}

\item {Cosmological Constant}

Inserting $w_\Lambda=-1=const.$ into (\ref{FFA3}) we obtain that $Q(a)=a^2$,
which leads to $f(a)=1$ and thus to $X(a)=2$.

\item{Quintessence and Phantom models with constant $w$}

In these constant-$w$ scenarios \cite{Ame10,Peebles03,phant} DE is
attributed to a homogeneous scalar field, with a suitable potential in order
to keep the EoS constant, which requires a form of fine tuning. Specifically,
the DE models with a canonical kinetic term of the scalar field lead to
$-1\le w$, while models of phantom DE ($w<-1$) require an  exotic
nature, namely a scalar field with negative kinetic energy, which could lead
to unstable quantum behavior \cite{Cline:2003gs}.
Substituting   $w(a)=w=const.$
into (\ref{FFA3}) we find
\begin{equation}
Q(a)=a^{-(1+3w)},
\end{equation}
and thus
\begin{equation}
\label{yLL}
X(a)=-(1+3w)a^{-3(1+w)} \;.
\end{equation}
 In other words, if we desire to construct a
Quintessence or Phantom look-alike Hubble expansion
(frequently used in cosmological studies), we need to
write $X(a)$ as in  (\ref{yLL}).

\item {Chevalier-Polarski-Linder DE}

We consider the Chevalier-Polarski-Linder parametrization
\cite{Chevallier:2001qy},
in which the dark energy EoS parameter is
defined as a first-order Taylor-expansion around the present
epoch:
\begin{equation}
w(a)=w_{0}+w_{1}(1-a).\label{cpldef}
\end{equation}
In this case we straightforwardly obtain
\begin{equation}
Q(a)=a^{-(1+3w_{0}+3w_{1})}{\rm exp}\left[ -3w_{1}(1-a)\right]
\end{equation}
and therefore
\begin{equation}
\label{yCPL}
X(a)=\frac{3w_{1}(a-1)-(1+3w_{0})}{a^{2}}\;Q(a)\;.
\end{equation}
 Similarly to the previous example,  if we want to build a CPL look-alike
Hubble expansion
in the context of Finsler-Randers geometry then the
corresponding functional form of $X(a)$ needs to obey (\ref{yCPL}).

\item  {Pseudo-Nambu Goldstone boson scenario}

In the Pseudo-Nambu Goldstone boson model \cite{Sorbo2007} the dark energy
EoS parameter is found with the aid of the potential $V(\phi)\propto [1+{\rm
cos}(\phi/p)]$ and it reads
\begin{equation}
w(a)=-1+(1+w_{0})a^{p},
\end{equation}
where $p$ is a free parameter of the model.
Based on this parametrization the basic ERF functions
are given by
\begin{equation}
Q(a)=a^{2}{\rm exp}\left[ -3\frac{1+w_{0}}{p}\left(a^{p}-1\right)\right]
\end{equation}
and
\begin{equation}
X(a)=\frac{2-3(1+w_{0})a^{p}}{a^{2}}\;Q(a)\;.
\end{equation}

\item  {$f(T)$ gravity}

Let us now give an example of how we can reconstruct the functional
forms of $X(a)$ and $Q(a)$ of the equivalent Finsler-Randers cosmology, in
the case of a modified gravitational model. As a specific case we choose the
$f(T)$ construction, which is based on the teleparallel equivalence of
General Relativity. In this formulation the gravitational information is
included in the torsion tensor and the corresponding torsion scalar $T$, and
one extends the Lagrangian considering arbitrary functions $f(T)$
\cite{Ben09,Ferraro:2006jd}.
Within such a framework the Hubble
function is written as
\begin{eqnarray}
\label{background1}
H^2= \frac{8\pi G}{3}\rho_m
-\frac{f(T)}{6}-2f_{T}H^2,
\end{eqnarray}
where $T=-6H^{2}$ is the torsion scalar and $f_{T}=\partial f(T)/\partial T$.
Based on the matter epoch, defining
$E^2_{FT}(a)=H^2(a)/H^2_0$ and using
$\rho_{m}=\rho_{m0}a^{-3}$, the above equation always becomes
\begin{equation}
\label{Mod1Ez}
E^2_{FT}(a)=\Omega_{m0}a^{-3}+\Omega_{F0} y(a)
\end{equation}
where $\Omega_{F0}=1-\Omega_{m0}$.
The function $y(a)$ is scaled to unity
at present time and is given by
\begin{equation}
\label{Mod2Ez}
y(a)=-\frac{1}{\Omega_{F0}}\left[ \frac{f(T)}{6H^{2}_{0}}+2f_{T}E^{2}(a)\right]
\;.
\end{equation}
Comparing relations (\ref{Mod1Ez}), (\ref{Mod2Ez}) with equation
(\ref{H2DE2}),
we find that
\begin{equation}
Q(a)=\frac{y(a)}{a^2}
\end{equation}
and
\begin{equation}
X(a)=\frac{\frac{dy}{d{\rm ln}a}-2y(a)}{a^{4}}.
\end{equation}

As an example we use the power-law model of Bengochea \& Ferraro
\cite{Ben09} with
\begin{equation}
\label{Mod3Ez}
f(T)=\alpha (-T)^{b}, \;\;\;\;\;\;\;
\alpha=(6H^{2}_{0})^{1-b}\frac{\Omega_{F0}}{2b-1},
\end{equation}
where $b$ is the free parameter of the model which has to be less than unity
in order to ensure a cosmic acceleration. Inserting (\ref{Mod3Ez}) into
(\ref{Mod2Ez}) we arrive at $y(a)=E^{2b}(a)=a^{2}Q(a)$. Obviously,
for $b=0$ the power-law $f(T)$ model reduces to the $\Lambda$CDM model, while
for  $b=1/2$ it reduces to the DGP   one \cite{Dvali2000}, which implies
that potentially we have a cosmological equivalence among the EFR, DGP and
$f(T)$ power-law gravity models. Note that, as we said in the Introduction,
the equivalence of DGP with Finsler-Randers cosmology was already found by
some of us in \cite{Bastav13}.

\end{itemize}

In summary, from the above analysis and the specific examples, it becomes
clear that the DE scenarios (including some modified gravity models)
that satisfy  (\ref{eq:Hgeneral}), can be seen as equivalent to the
geometrical EFR cosmological model.

 Finally, in a forthcoming publication we attempt
to physically derive the precise functional form of $X(a)$, as well
as to provide a full perturbation analysis, which can
be used in order to distinguish
the Finser-Randers scenario from other DE and modified gravity models
\cite{usfuture}. 

\section{Conclusions}

In the present work we investigated an extended form of Finsler-Randers
cosmology, and we showed that it can mimic any   non-interacting
dark-energy scenario, as well
as modified gravity models, at the background level. In particular, we
started from a small deviation from the quadraticity of the Riemannian
geometry, and we extracted the modified Friedmann equation that determines
the universe evolution.

The effect of the Finsler-Randers modification is to produce correction terms
to the Friedmann equation, that can lead to a large class of scale-factor
evolution, including the quintessence and phantom regimes, the phantom-divide
crossing from both sides, etc. As we showed, for a given dark-energy
equation-of-state parameter we can reconstruct the corresponding functions of
the Finsler-Randers space that indeed give rise to such a behavior, and vice
versa. Therefore, the present work is a completion of the previous works of
some of us \cite{Stavrin,Bastav13}, where we had showed the equivalence of
Finsler-Randers cosmology with particular modified gravitational models as
the DGP one, since we now   show  that
the extended Finsler-Randers cosmology can
  resemble a large class  of cosmological scenarios.

In this respect, the non-trivial universe evolution, and especially its
accelerated phase either during inflation or at late times, is not
attributed to a new scalar field, or to gravitational modification, but it
arises from the  modification of the geometry itself. In particular, even a
very small non-quadraticity of the Finsler-Randers geometry, in which the
local structure of General Relativity is modified and the curvature theory
is extended, can lead to significant implications to the cosmological
evolution. One should still provide an explanation for the origin of
the Finsler-Randers geometry itself, and the small departure from the
Riemann one. Although there are indications that this must be related to
quantum gravity effects
\cite{Barcelo:2005fc,Mavromatos:2010ar,Germani:2011bc,Romero:2009qs,
Cohen:2006ky,Leigh:2011au,Skakala:2010hw,Kostelecky:2011qz},
 this issue lies beyond the scope of the present work and it is left for  a
future investigation.

We close this work by making two comments. The first is that, as we
discussed in the text, our analysis is valid at intermediate and late times,
including the radiation era, where all the energy conservations hold as
usual. The second is that the above equivalence between
Finsler-Randers geometry and dark energy and modified gravity models, has
been obtained at the background level, that is demanding the same
scale-factor evolution. However, a necessary step is to proceed to a
detailed analysis of the cosmological perturbations, and see whether the
aforementioned equivalence breaks, which would allow  to distinguish
between the various scenarios (this was indeed the case in the
equivalence of the simple Finsler-Randers geometry with the DGP model
\cite{Bastav13}), or whether it is maintained, in which case the degeneracy
of the above constructions would be complete. This complicated and detailed
investigation is in progress \cite{usfuture}.

\begin{acknowledgments}
The authors would like to thank an unknown referee for his valuable comments
and suggestions.
SB acknowledges support by the Research
Center for Astronomy of the Academy of Athens
in the context of the program  ``{\it Tracing the Cosmic Acceleration}''.
The research of ENS is implemented within the framework of the Action
``Supporting Postdoctoral Researchers'' of the Operational Program
``Education and Lifelong Learning'' (Actions Beneficiary: General Secretariat
for Research and Technology), and is co-financed by the European Social Fund
(ESF) and the Greek State. PS receives partial support from the ``Special
Accounts for Research Grants''  of the University of
Athens.

\end{acknowledgments}

\appendix*

\section {The 1+3 covariant formalism}

We briefly summarize the 1+3 covariant formalism as developed by J.Ehlers and G.F.R.Ellis \cite{Relcos}. The notation in this Appendix is for the Riemannian limit that will serve as guidance through our calculations to the Finslerian case. The covariant approach employs a time-like vector field $u^a$ with $u_au^a=1$. With respect to this normalized vector field we split spacetime to time and space. The 1+3 split is a particular case of the tetrad formalism where the $u^a$ congruence represents the frame of comoving observers. With respect to the 4-velocity $u^a$ we can decompose all tensors to their irreducible parts.  In particular, using the projection 2nd-rank tensor $h_{ab}=g_{ab}-u_au_b$ we can covariantly define the time derivative and the spatial gradient of an arbitrary tensor field, namely
\begin{eqnarray}
\dot S_{ab..}^{\;\;\;\;cd..}&=&u^e\nabla_eS_{ab..}^{\;\;\;\;cd..}\\
{\rm D}_eS_{ab..}^{\;\;\;cd..}&=&h_e^{\;\;s}h_a^{\;\;f}h_b^{\;\;p}h_q^{\;\;c}h_r^{\;\;d}...\nabla_sS_{fp..}^{\;\;\;\;qr..}.
\end{eqnarray}
Instead of writing the metric to a particular coordinate system the geometry as measured by the $u^a$ family of observers is described by the irreducible parts of the following tensor field
\begin{equation}
{\rm D}_bu_a={1\over3}\Theta h_{ab}+\sigma_{ab}+\omega_{ab},
\end{equation}
where we define the kinematic quantities: the expansion $\Theta={\rm D}^au_a$, the shear $\sigma_{ab}={\rm D}_{\langle
b}u_{a\rangle}$ and the vorticity $\omega_{ab}={\rm D}_{[
b}u_{a]}$. The projective symmetric and trace free part is defined as
\begin{equation}
X_{\langle
ab\rangle}=h^c_{\;\;(a}h^d_{\;\;b)}X_{cd}-{1\over3}X_{cd}h^{cd}h_{ab},
\end{equation}
where indices in squared brackets is the symmetrised part.

In case of a shear and vorticity free expanding congruence of geodesics  the
evolution of the deviation vector that connects nearby observers is
$\dot\xi_a={1\over3}\Theta\xi_a$ \cite{Relcos}. Taking the time derivative of
the previous expression, and substituting to the deviation of geodesics
$\ddot\xi_a+R_{acbd}u^cu^d\xi^b=0$, gives back the Raychaudhuri's equation
\begin{equation}
\dot{\Theta}+{1\over3}\Theta^2=-R_{ab}u^au^b.\label{Rexp}
\end{equation}
The energy momentum tensor of pressureless matter is $T_{ab}=\rho u_au_b$,
and the contracted Einstein's field equations along the observers 4-velocity
give back the auxiliary expression, $R_{ab}u^au^b={1\over2}\rho$. Moreover,
an important geometric entity is the characteristic length scale of the
expanding 3D cross-section, namely the scale factor $a$ given by $dV\propto
a^3$. The reader should notice that the scale factor is covariantly defined,
in contrast to the metric based approach where it is introduced through a
particular coordinate system.  Thus, for the expansion we obtain
$\Theta={(dV)\dot{}/dV}=3\dot a/a$, and therefore we can rewrite relation
(\ref{Rexp}) in the  form
\begin{equation}
3{\ddot a\over a}=-{1\over2}\rho.\label{ddota}
\end{equation}
The physical requirement of pressureless matter implies that for a
conservative system the matter energy density scales with the
volume element, that is $\rho dV=const$. Furthermore, using the definition
for the scale factor we acquire $(\rho a^3)\dot{}=0$ (alternatively one may
decompose the energy momentum conservation law $\nabla^bT_{ab}=0$ to its'
irreducible parts \cite{Relcos}). The latter together with relation
(\ref{ddota}) fully determines the evolution of the dust-like medium (for
further details see for example \cite{ellispr}).

\end{document}